\documentclass[aps, prx, superscriptaddress, reprint , twocolumn, floatfix, preprintnumbers]{revtex4-2}
\usepackage{latexsym, amsmath, amsfonts, amssymb}

\usepackage{graphicx}

\usepackage{comment}

\usepackage{hyperref}
\hypersetup{colorlinks = true, linkcolor = blue, anchorcolor = blue, citecolor = blue, filecolor = blue, urlcolor = blue}
\usepackage{bbm}
\usepackage{mathrsfs}
\usepackage{array} 
\usepackage{xcolor}

\usepackage{tikz}
\usepackage{tikz-3dplot}
\usetikzlibrary{fadings,shadings}
\usetikzlibrary{decorations.pathmorphing}
\usepackage{enumitem}

\setlist[itemize]{leftmargin=1.25em}
\setlist[enumerate]{leftmargin=1.25em}

\newcommand{\ds}{\displaystyle}

\newcommand{\be}{\begin{equation}} \newcommand{\ee}{\end{equation}}
\newcommand{\bea}{\begin{equation} \begin{aligned}} \newcommand{\eea}{\end{aligned} \end{equation}}

\newcommand{\bQ}{\mathbb{Q}}

\newcommand{\bZ}{\mathbb{Z}}

\newcommand{\unit}{\mathbbm{1}}

\newcommand{\calM}{\mathscr{M}}

\newcommand{\bx}{\mathbf{x}}
\newcommand{\by}{\mathbf{y}}
\newcommand{\bz}{\mathbf{z}}

\DeclareMathOperator{\Tr}{Tr}

\begin{document}

\title{'t Hooft Anomalies and Defect Conformal Manifolds:\\[0.5em] Topological Signatures from Modulated Effective Actions}

\newcommand{\OXFORD}{\affiliation{Mathematical Institute, University
of Oxford, Woodstock Road, Oxford, OX2 6GG, United Kingdom}}

\author{Christian Copetti}
\OXFORD

\begin{abstract}
Symmetry breaking of continuous symmetries by extended dynamical defects entails the existence of defect families, which form conformal manifolds in a critical setup. In the presence of bulk 't Hooft anomalies, defects are in fact required to break the symmetry \cite{Jensen:2017eof,Thorngren:2020yht}: defect conformal manifolds are anomaly-enforced. We show that, by coupling the system to a modulated deformation parameter, the geometric structure of the conformal manifold is sensitive to the 't Hooft anomaly. This leads to measurable effects in the presence of a boundary/defect: in (1+1)d the anomaly predicts a quantized boundary charge pumping, while in higher dimensions it gives rise to non-dissipative boundary Hall currents.
\end{abstract}

\maketitle

\section{Introduction}
Symmetry breaking induced by a dynamical defect $D$ has attracted increasing attention in recent years, particularly in the context of conformal theories (DCFTs). It was observed in \cite{Padayasi:2021sik} and further developed in subsequent works \cite{Drukker:2022pxk,Cuomo:2021cnb,Herzog:2023dop,Antinucci:2024izg} that symmetry breaking by a defect is encoded in the physics of the so-called \emph{tilt} operator, defined via\footnote{Throughout this work, we use $x, y, \ldots$ to denote bulk coordinates and $\bx, \by, \ldots$ for defect coordinates.}
\be
\partial_\mu J^\mu_a(x) = t_a(\bx) \, \delta(\Sigma) \, , \ \
\ee
where $\delta(\Sigma)$ is a delta function supported on the defect worldvolume $\Sigma$. In a conformal setting, the tilt operator $t_a(\bx)$ has a protected scaling dimension $\Delta_t = p$, where $p$ is dimension of $\Sigma$, making it exactly marginal. Notably, the existence of the tilt operator is not limited to conformal theories:
even in the absence of conformal symmetry, the breaking of a continuous symmetry gives rise to a \emph{continuous family} of defects, related by the action of the bulk symmetry. 
Even in such cases, we will (improperly) refer to these families as conformal manifolds and denote them by $\calM$. These are generically homogeneous spaces\footnote{Exceptions exist for interfaces in (1+1)d \cite{Antinucci:2025uvj}; for example, see defects in the Ising CFT \cite{Oshikawa:1996dj}.} of the form $\calM = G/H$, where $H$ is the subgroup of $G$ left unbroken by the defect.
The bulk symmetry operators $U_h$, $h \in G$, act on $\calM$ by left multiplication. This is induces by encircling of the defect by the topological symmetry operator $U_h$. see Figure \ref{fig:action}. In this paper, we mainly focus on the breaking of the full $G$ symmetry by the defect, and denote the coordinate on $\calM$ by $g = e^{i \sigma}$.
\begin{figure}
\begin{tikzpicture}
    \draw[color= cyan, line width =2] (0,0) node[left,black] {$D_g$} -- (0,3);
 \begin{scope}[shift = {(0,3)}]   
   \def\rx{1}    
  \def\ry{0.4}  
  \def\h{3}     

  \draw (0,0) ellipse ({\rx} and {\ry});

  \draw (-\rx,0) -- (-\rx,-\h);
  \draw ( \rx,0) -- ( \rx,-\h);

  \draw (0,-\h) ellipse ({\rx} and {\ry});

  \fill[blue!70!white,opacity=0.5]
    (-\rx,-\h)                         
      arc (-180:0: 1 and 0.4)        
      -- (\rx,0)                  
      arc (0:-180: 1 and 0.4)        
      -- cycle;            
      \node[left] at (-\rx,-\h) {$U_h$};

      \node[right] at (\rx,-\h/2) {$\quad =$};
\end{scope}

\begin{scope}[shift={(3,0)}]
        \draw[color= cyan, line width =2] (0,0) node[left,black] {$D_{hg}$} -- (0,3);
\end{scope}
\end{tikzpicture}
    \caption{The action of a bulk symmetry operator $U_h$ on a symmetry-breaking defect $D_g$.}
    \label{fig:action}
\end{figure}

It is widely appreciated that the symmetry of a bulk system is described by more data than the choice of a symmetry group. In particular, the bulk symmetry $G$ might be anomalous:
\be
Z[A +d_A \lambda] = \exp\left( i \int \alpha(A, \lambda) \right) Z[A] \, .
\ee
't Hooft anomalies are routinely characterized via anomaly inflow by instead considering a $d+1$-dimensional invertible theory $\omega(A) \in \textbf{SPT}^{d+1}_G$, satisfying:
\be
\delta_\lambda \omega(A) = d \alpha(A, \lambda) \, .
\ee
It is well known that in the presence of nontrivial anomalies, boundary conditions cannot preserve the full symmetry \cite{Jensen:2017eof,Thorngren:2020yht}. For continuous symmetries, this is a consequence of the non-vanishing contact terms in the correlation functions of the associated current $J^\mu$.
A similar conclusion holds for higher co-dimension defects \cite{Copetti:2024onh}, provided  the bulk anomaly is sourced by the defect background\footnote{A typical example is the presence of background magnetic fluxes, commonly encountered in supersymmetric backgrounds.}. We will refer to this phenomenon as \emph{anomaly-enforced symmetry breaking}. It is natural to ask whether the symmetry-breaking pattern in the presence/absence of the anomaly can differ, and how the bulk anomaly might be detected by defect observables.

For conformal defects, the natural objects to study are the correlation functions of the tilt operators:
\be
G_{a_1,...,a_n}(\bx_1, ..., \bx_n) = \langle t_{a_1}(\bx_1) ... t_{a_n}(\bx_n) \rangle \, .
\ee
The 2-pt function $G_{ab}(\bx_1,\bx_2)$ encodes the Zamolodchikov metric on the defect conformal manifold \cite{Zamolodchikov:1986gt}, while the 3-pt function is known to vanish (modulo contact terms): this is a requirement for the absence of an anomalous dimension. 
Geometric information about the conformal manifold can also be extracted from contact terms in the OPE\cite{Seiberg:1988pf,Kutasov:1988xb,Drukker:2022pxk,Drukker:2022txy,Herzog:2023dop}:
\be
t_a(\bx) t_b(\by) = \frac{g_{ab}}{(\bx - \by)^{2p}} + \Gamma_{a b}^c \, t_c(\by) \, \delta^{p}(\bx - \by) + ... \, ,
\ee
where $\Gamma_{ab}^c$ gives a connection on $\calM$. We will discuss how, in the presence of the bulk anomaly, further universal data appear in this OPE.

For non-conformal defects, the natural quantities of interest are instead response functions. These can be defined by coupling the bulk-defect system to external sources (gauge fields) and studying the resulting partition function $Z_D(A,\sigma)$, where $A$ is an external gauge field for the (broken) symmetry.
When the defect is placed on a sphere $S^p$, we can construct out of this quantity a
\emph{Defect Free Energy}:\footnote{For a boundary condition, one would divide by $|Z[A]|^{1/2}$ instead.}
\be
F_D[A,\sigma] = - \log \left( \frac{Z_D[A, \sigma]}{ \big\vert  Z[A] \big\vert} \right) \, .
\ee
This has been widely used in the study of defect monotonicity theorems \cite{Friedan:2003yc,Jensen:2015swa,Kobayashi:2018lil,Cuomo:2021rkm,Gaiotto:2014gha,Giombi:2020rmc,Wang:2020xkc}. The defect free energy is generally not a local functional of $A, \sigma$ and its derivatives.

In this Letter, we leverage the bulk 't Hooft anomaly to constrain universal terms in the response functions stemming from $F_{D}$. 
The interplay between anomalies and effective actions has, in recent years, led to several remarkable results \cite{Son:2009tf,Landsteiner:2011cp,Jensen:2012kj,Jensen:2013kka,Jensen:2013rga,DiPietro:2014bca,Assel:2015nca,Ohmori:2021dzb}, including striking applications to non-dissipative transport in condensed matter systems (see e.g. \cite{Landsteiner:2016led} for a review). Our work extends these ideas by incorporating symmetry-breaking defects. The main results are:
\begin{enumerate}
    \item \emph{Charge pumps, transport, and the topology of $\calM$.} The defect effective action can be used to derive non-dissipative defect currents induced by the anomaly. For (1+1)d boundary conditions, this takes the form of a Thouless charge pump; for (3+1)d systems, it manifests as a defect analogue of the Quantum Hall and Chiral Magnetic effects \cite{Fukushima:2008xe}, resulting in the pumping of an integer Quantum Hall phase. These SPT-pumping phenomena connect to the global topology of the defect conformal manifold.
     \item \emph{Anomalies in the space of defect couplings.} We argue that these effects can be understood as anomalies in the space of \emph{defect} couplings, generalizing the framework of \cite{Cordova:2019jnf,Cordova:2019jnf}. Furthermore, if the bulk-defect system exhibits a \emph{Defect Anomaly} \cite{Antinucci:2024izg}, we propose that it can be matched on a symmetry-breaking defect through analogous mechanisms.
\end{enumerate}
We further comment on the mechanism with which the bulk anomaly sources novel contact terms in the tilt OPE, and how these are related to the anomaly by modulated WZW terms.
In this work, we focus on \emph{continuous}, \emph{group-like} symmetries and their interplay with defect dynamics. Generalizations to spacetime and non-invertible/higher-form symmetries will be explored in future work.

\vspace{2mm}
\noindent\emph{Note Added.} As this work was nearing completion, the papers  \cite{Choi:2025ebk,Wen:2025xka}, in which similar effects are studied in (1+1)d from the perspective of boundary Berry Phases, appeared on the arXiv. The approach developed here is complementary to \cite{Choi:2025ebk,Wen:2025xka}, and the conclusions reached in (1+1)d are compatible. Before v2 of this work was finalized, related ideas were also discussed in \cite{Drukker:2025dfm}.

\section{Symmetric and Modulated Defects}
We begin by briefly reviewing how symmetric defects can be coupled to bulk gauge fields, and how to extend this construction to symmetry-breaking defects. This will require us to consider a new type of defect, in which the symmetry breaking parameter $g$ is taken to vary continuously over the defect's worldvolume coordinates. We will call this a ``modulated" defect. The main advantage of this construction is that, unlike standard symmetry-breaking defects, modulated defects can be coupled consistently to background gauge fields $A$ for the broken symmetry, allowing for a clear discussion of the interplay between anomalies and defect observables.

\begin{figure}
    \centering
    \begin{tikzpicture}
     \draw[color=blue!70!white,fill=blue!70!white, opacity=0.5] (1,0.5) -- (1,2.5) -- (-1,2.5) -- (-1,0.5) -- cycle;
        \draw[color=white!70!cyan, fill=white!70!cyan,opacity=0.8] (0,0) node[left, color=black, opacity=1] {$D$} -- (2,1) -- (2,3) -- (0,2) -- cycle;
        \draw[color=blue!50!white,fill=blue!50!white,opacity=0.8] (1,0.5) -- (1,2.5) -- (3,2.5) -- (3,0.5) -- cycle;
        \draw[thick] (1,0.5) -- (1,2.5);
        \node at (-0.75, 2.25) {$U_g$};
        \node at (-1.5,1.5) {(a)};
        \node[right] at (1,3.5) {$\ds \int_{\text{\tiny $U_g \cap D$}} \star j$};
        \draw[->,looseness=2] (1.1,3.2) to[out = 200, in =90] (1,2.55);
    \end{tikzpicture} \\[2em]
    \begin{tikzpicture}
      \draw[color=white!70!cyan, fill=white!70!cyan,opacity=0.5] (0,0) node[left, color=black, opacity=1] {$D_\sigma$} -- (2,1) -- (2,3) -- (0,2) -- cycle;  
      \draw[->] (2.5,1.5) -- (3.5,1.5);
      \node[above] at (3,1.5) {$\delta_{\lambda(\bx)}$};
   \begin{scope}[shift={(4,0)}]
   \def\N{100}        
  \def\A{0.1}        
  \def\freq{4}       

  \fill[white!70!cyan, opacity=0.5]
    plot[domain=0:1, samples=\N, variable=\t]
      ({\A*sin(360*\freq*\t)}, {2*\t}) --       
    plot[domain=1:0, samples=\N, variable=\t]
      ({2 + \A*sin(360*\freq*\t)}, {1 + 2*\t}) -- 
    cycle;
    \node[left] at (0,0) {$D_{\lambda(\bx)}$};

      \foreach \i in {0,...,\N} {
    \pgfmathsetmacro{\t}{\i/\N}
    \pgfmathsetmacro{\xleft}{\A*sin(360*\freq*\t)}
    \pgfmathsetmacro{\xright}{2 + \A*sin(360*\freq*\t)}
    \pgfmathsetmacro{\y}{2*\t}
    \pgfmathsetmacro{\shadeval}{abs(sin(360*\freq*\t))}
    \pgfmathsetmacro{\opacity}{0.05 + 0.15*\shadeval}
    \draw[cyan!50!black, line width=0.5pt, opacity=\opacity]
      (\xleft,\y) -- (\xright,{1 + \y});
  }

      \end{scope}

      \node at (-0.5,1.5) {(b)};
    \end{tikzpicture}
    \caption{(a) A symmetric defect can be crossed topologically by the bulk symmetry generators via an improvement term. (b) A symmetry-breaking defect is mapped into a \emph{modulated} defect by a background gauge transformation.}
    \label{fig: symandmod}
\end{figure}

\vspace{2mm}\noindent {\bf Symmetric Defects}  The discussion follows closely \cite{Antinucci:2024izg}. Consider a continuous 0-form symmetry $G$, generated by the topological defects:
\be
U_g = \exp\left(i \sigma^a \int \star J_a \right) \, , \quad g = e^{i \sigma^a T_a}
\ee
A defect $D$ is $G$-symmetric if and only if:
\be
U_g \, D = D \, U_g \, .
\ee
Equivalently, the $U_g$ defect can traverse $D$ topologically, as shown in Figure~\ref{fig: symandmod}. This is reflected in an improvement term for the bulk current, defined by a defect-localized current $j_a$:
\be \label{eq: trivialtilt}
t_a(\bx) = - d \star_{\Sigma} \, j_a (\bx) \, .
\ee
In the presence of a defect current $j_a$, the bulk-defect system admits coupling to a background gauge field $A$, via:
\be
S[A] = 
S_0 + i\int \Tr A \wedge \star J + i \int_{\Sigma} \Tr A_D \wedge \star j \, ,
\ee
where $A_{D}$ denotes the pullback of $A$ on the defect's worldvolume. In the absence of bulk anomalies, the resulting defect partition function is 
invariant under background gauge transformations $A \to A^{g} = g^{} A g^{-1} - d g\,  g^{-1}$, thanks to \eqref{eq: trivialtilt}. 
Furthermore, the presence of a background gauge field allows us to discuss the notion of \emph{defect anomalies}, that is, 't Hooft anomalies localized on the defect's worldvolume \cite{Antinucci:2024izg}.
These are encoded in a nontrivial transformation of the defect partition function under gauge transformations:
\be
Z_D[A + d_A \lambda, \mathbf{B}] = \exp\left( i \int_{\Sigma} \alpha_D(A, \mathbf{B}, \lambda) \right) Z_D[A, \mathbf{B}] \, ,
\ee
where $\mathbf{B}$ denote gauge fields for defect symmetries $\mathbf{K}$, i.e. symmetries which are decoupled form bulk degrees of freedom.
Similar to standard 't Hooft anomalies, defect anomalies  are classified by invertible phases $\omega_D(A, \mathbf{B}) \in \textbf{SPT}^{p+1}_{G \times \mathbf{K}}$. 

\vspace{2mm} \noindent {\bf Modulated Defects} We now move on to discuss symmetry-breaking defects. If the bulk symmetry is broken, the bulk current $J^\mu_a$ fails to be conserved on the defect's worldvolume and the (integrated) tilt operator $t_a(\bx)$ is non-trivial. 
At first glance, this seems to prevent us coupling the system to a background gauge field $A$ for the symmetry: under an infinitesimal gauge transformation, the bulk minimal coupling shifts the defect action by:
\be
\delta_\lambda S = i \int_{D} \Tr \lambda(\bx)  \, t(\bx) + ...  \, , 
\ee
where the dots denote higher order terms in $\sigma$. The new defect is described by a position-dependent deformation along its conformal manifold $\calM$, as illustrated in Figure~\ref{fig: symandmod}. 
This issue can be resolved by considering a \emph{modulated} defect from the get-go instead. That is, we promote the symmetry breaking parameter $g$ to a function of the defect coordinate $g(\bx)$ instead, and postulate the background transformations:
\be
(A, g) \longrightarrow (A^h, h g) \, .
\ee
This can be understood as restoring the defect symmetry $G$ by the coupling to a spurionic NLSM with target $G$. 
The invariance can be seen very explicitly by working at linear order near the identity (in both $\lambda$ and $\sigma$): $g \sim \unit + i \sigma$ whence $\delta_\lambda \sigma = -\lambda$ and the spurionic transformation exactly cancels the bulk contribution stemming from the tilt operator.

The modulated free energy $F_D[A,\sigma]$ is promoted to a (background) gauge invariant functional of the gauge field $A$ and the NLSM field $g$. This approach has been used in \cite{Cuomo:2023qvp} to describe spontaneous symmetry breaking on defects.\footnote{Related effective actions, involving the \emph{displacement} of the defect's position, have recently been studied in \cite{Gabai:2025zcs}.} In this work we will focus in describing properties of the free energy which descend from bulk 't Hooft anomalies.

\section{Anomalous Modulated Effective Actions}
Having discussed how to consistently couple background gauge fields to symmetry-breaking defects via modulation, we now incorporate perturbative bulk 't Hooft anomalies in our analysis and examine their impact on the structure of $Z_D$.
We will start by discussing the simpler case of boundary conditions, and explain how some of the methods describe also generalize to defects of various types. The mathematical framework underlying these results essentially coincides with the Symmetry-Breaking Exact Sequence of \cite{Debray:2023ior}. We will focus on a pedagogical presentation, which will however suffice for our applications.

Recall that an 't Hooft anomaly is encoded in the nontrivial transformation of the partition function in the presence of background gauge fields:
\be
Z[A + d_A \lambda] = e^{ i \int \alpha(A,\lambda)} Z[A] \, .
\ee
The anomaly $\alpha$ satisfies the Wess–Zumino consistency condition:
\bea
&\text{WZ}(A,\lambda_1,\lambda_2)  \\
&= \delta_{\lambda_1} \alpha(A,\lambda_2) - \delta_{\lambda_2} \alpha(A, \lambda_1) - \alpha(A,[\lambda_1, \lambda_2]) = 0 \, .
\eea
A nontrivial anomaly cannot be eliminated by a local counterterm, i.e., $\alpha(A,\lambda) \neq d \beta(A,\lambda)$. 
In the absence of boundaries, the Wess-Zumino condition is can be violated by total derivatives:
\be
\text{WZ}[A,\lambda_1, \lambda_2] = d \beta(\lambda_1, \lambda_2, A) \, .
\ee
These can be dropped in the routine analysis. On the other hand, in the presence of boundaries, it was observed in \cite{Jensen:2017eof} that
such violations become physically relevant and 
cannot be remedied by local (boundary) counterterms, leading to an \emph{anomaly-enforced symmetry breaking}.
Once symmetry-breaking is assumed, however, mileage is gained by explicitly solving these equations in the presence of a modulated symmetry-breaking parameter $\sigma(\bx)$.
Naturally, the partition function in the presence of a symmetry-breaking boundary can itself transform projectively under gauge transformations:\footnote{The explicit form of $\kappa(\lambda,\sigma)$ can be inferred from the Baker-Campbell-Haussdorf formula applied to the right $G$ action.}
\bea
&Z_B[A + d_A \lambda, \sigma - \kappa(\lambda,\sigma) ] \\
&= \exp\left(i \int \alpha(A,\lambda) - i \int_B \alpha_B(A,\sigma;\lambda) \right) Z_B[A, \sigma]
\eea
Physically, we can interpret $\alpha_B(A,\sigma;\lambda)$ as the contribution to the defect effective action stemming from passing from a standard symmetry-breaking defect to a modulated one. The consistency condition now reads:
\bea
&\beta(\lambda_1,\lambda_2,A) \\
&= \delta_{\lambda_2} \alpha_{B}(A,\sigma;\lambda_1) - \delta_{\lambda_1} \alpha_{B}(A,\sigma;\lambda_2) - \alpha_B (A, \sigma; [\lambda_1, \lambda_2]) \, .
\eea
Solving this equation determines $\alpha_B$ up to an homogeneous solution of the boundary consistency condition $\nu_B$. The non-homogenous term is a manifestation of the bulk anomaly on the boundary, which we will focus on throughout much of the presentation. On the other hand, we will refer to the homogeneous solutions $\nu_B$ as a ``defect anomaly".
While the former is universal, in the sense that it is shared by all boundary conditions supporting the same bulk anomaly (and having the same symmetry-breaking pattern), the latter depends on the choice of boundary/defect we consider. In the presence of a nontrivial bulk anomaly, the defect anomaly is not a well-defined concept. However, considering two defects $D_1$ and $D_2$ with the same bulk, the relative defect anomaly becomes well defined.

\begin{figure}
    \centering
    \begin{tikzpicture}
 \draw[color=white!70!cyan, fill=white!70!cyan,opacity=0.5] (0,0) -- (2,1) -- (2,3) -- (0,2) -- cycle;  
     \node[left] at (-0.25,1.5 ) {$D$};
    \draw[->] (2.25, 1.5 ) -- (3.25,1.5); \node[above] at (2.75,1.5) {$\delta_{\lambda_1}$};
    \draw[->] (1, 0.25 ) -- (1,-0.75); \node[left] at (1,-0.25) {$\delta_{\lambda_2}$};

    \begin{scope}[shift={(3.5,0)}]
     \def\N{100}        
  \def\A{0.1}        
  \def\freq{4}       

  \fill[white!70!cyan, opacity=0.5]
    plot[domain=0:1, samples=\N, variable=\t]
      ({\A*sin(360*\freq*\t)}, {2*\t}) --       
    plot[domain=1:0, samples=\N, variable=\t]
      ({2 + \A*sin(360*\freq*\t)}, {1 + 2*\t}) -- 
    cycle;
    \node[right] at (2.25,1.5) {$D_{\lambda_1(\bx)}$};

      \foreach \i in {0,...,\N} {
    \pgfmathsetmacro{\t}{\i/\N}
    \pgfmathsetmacro{\xleft}{\A*sin(360*\freq*\t)}
    \pgfmathsetmacro{\xright}{2 + \A*sin(360*\freq*\t)}
    \pgfmathsetmacro{\y}{2*\t}
    \pgfmathsetmacro{\shadeval}{abs(sin(360*\freq*\t))}
    \pgfmathsetmacro{\opacity}{0.05 + 0.15*\shadeval}
    \draw[cyan!50!black, line width=0.5pt, opacity=\opacity]
      (\xleft,\y) -- (\xright,{1 + \y});
  }
           \draw[->] (1, 0.25 ) -- (1,-0.75); \node[left] at (1,-0.25) {$\delta_{\lambda_2}$};
    \end{scope}

     \begin{scope}[shift={(0,-3.5)}]
      \def\N{100}        
  \def\A{0.1}        
  \def\freq{4}       

  \fill[white!70!cyan, opacity=0.5]
    plot[domain=0:1, samples=\N, variable=\t]
      ({\A*sin(360*\freq*\t)}, {2*\t}) --       
    plot[domain=1:0, samples=\N, variable=\t]
      ({2 + \A*sin(360*\freq*\t)}, {1 + 2*\t}) -- 
    cycle;
    \node[left] at (-0.25,1.5) {$D_{\lambda_2(\bx)}$};

      \foreach \i in {0,...,\N} {
    \pgfmathsetmacro{\t}{\i/\N}
    \pgfmathsetmacro{\xleft}{\A*sin(360*\freq*\t)}
    \pgfmathsetmacro{\xright}{2 + \A*sin(360*\freq*\t)}
    \pgfmathsetmacro{\y}{2*\t}
    \pgfmathsetmacro{\shadeval}{abs(sin(360*\freq*\t))}
    \pgfmathsetmacro{\opacity}{0.05 + 0.15*\shadeval}
    \draw[cyan!50!black, line width=0.5pt, opacity=\opacity]
      (\xleft,\y) -- (\xright,{1 + \y});
  }
        \draw[->] (2.25, 1.5 ) -- (3.25,1.5); \node[above] at (2.75,1.5) {$\delta_{\lambda_1}$};

    \end{scope} 
    \begin{scope}[shift={(3.5,-3.5)}]

      \def\N{100}        
  \def\A{0.1}        
  \def\freq{4}       

  \fill[white!70!cyan, opacity=0.5]
    plot[domain=0:1, samples=\N, variable=\t]
      ({\A*sin(360*\freq*\t)}, {2*\t}) --       
    plot[domain=1:0, samples=\N, variable=\t]
      ({2 + \A*sin(360*\freq*\t)}, {1 + 2*\t}) -- 
    cycle;
    \node[right] at (2.25,1.5) {$D_{\lambda_1(\bx)+\lambda_2(\bx)}$};

      \foreach \i in {0,...,\N} {
    \pgfmathsetmacro{\t}{\i/\N}
    \pgfmathsetmacro{\xleft}{\A*sin(360*\freq*\t)}
    \pgfmathsetmacro{\xright}{2 + \A*sin(360*\freq*\t)}
    \pgfmathsetmacro{\y}{2*\t}
    \pgfmathsetmacro{\shadeval}{abs(sin(360*\freq*\t))}
    \pgfmathsetmacro{\opacity}{0.05 + 0.15*\shadeval}
    \draw[cyan!50!black, line width=0.5pt, opacity=\opacity]
      (\xleft,\y) -- (\xright,{1 + \y});
  }
    \end{scope}
    \end{tikzpicture}
    \caption{The Wess–Zumino consistency condition in the presence of a symmetry-breaking defect $D$. Straight defects denote an unmodulated symmetry-breaking, while wiggly ones correspond to modulated defects.}
    \label{fig: modulated}
\end{figure}
For a defect $D$ of dimension $p$, we can derive similar constraints in the presence of background fluxes. To see this we excise a tubular neighbourhood $T\Sigma \simeq \Sigma \times S^{d-p-1}$ from spacetime. The defect is then defined by imposing a suitable boundary conditions on this geometry. Reducing the system on the $S^{d-p-1}$ we find again the same consistency conditions involving the reduced anomaly.
For a conformal defect, this corresponds to the well-known conformal mapping of the flat defect onto $\text{AdS}_{p+1} \times S^{d-p-1}$.

\vspace{2mm} \noindent \textbf{Inflow perspective} These equations are most easily understood with the help of anomaly-inflow. In the standard treatment, an anomalous system (with anomaly $\alpha$) can be realized in an anomaly-free manner on the boundary of an invertible topological theory $\omega(A) \in \textbf{SPT}_G^{d+1}$. Alternatively, we can think of $\omega(A)$ as a book-keeping device to describe the anomaly $\alpha$, via:
\be
\delta_\lambda \omega(A) = d \alpha(A, \lambda) \, .
\ee
Notice that, since $\omega$ is a local functional of $A$, the WZ consistency condition is automatically satisfied.

In the presence of a boundary condition, the inflow theory itself must admit an invertible topological boundary, see Figure \ref{fig: inflow} matching the anomaly.
This is impossible by the very definition of an SPT phase, but can be circumvented if we decorate the topological boundary by a spurionic symmetry breaking field $g$, with an action $\gamma(A,g)$. In our setup, this should be understood as a continuation of the modulated field $g(\bx)$ into the inflow bulk. Anomaly matching requires that:
\be
\alpha(A, \lambda) + \delta_\lambda \gamma(A,g) = d \alpha_B(A,g;\lambda) \, .
\ee
Again, the consistency condition in the presence of a boundary is automatically satisfied and our task is reduced to the description of a suitable $\gamma$; similarly, homogeneous solutions to $\delta_\lambda \gamma = \nu_B$ describe boundary/defect anomalies. 
For a perturbative anomaly, a solution to this problem is readily at hand. Consider the (left) gauge invariant combination:
\be
A_g = g^{-1} A g + g^{-1} d g \, ,
\ee
clearly, the anomaly $\omega(A_g)$ is exactly gauge invariant. However, as $A$ and $A_g$ are gauge equivalent, their anomalies differ by a surface term:
\be
\omega(A_g) - \omega(A) = d \gamma(A,g) \, .
\ee
For a perturbative anomaly this amounts to (a trivialization of) the appropriate WZW term (a B-field). Physically, we can interpret our construction as considering a symmetric interface between our theory and a slab of $G$ WZW model with the correct anomaly, supplemented with a Dirichlet boundary condition for the WZW fields. 

\vspace{2mm} \noindent \textbf{Partial breaking} Let us close with a short comment on partial breaking of the symmetry on the defect.  An anomaly-free subgroup is defined by a group homomorphism $\varphi : G_0 \to G$ which trivializes the anomaly \cite{Thorngren:2020yht}:
\be
\varphi^*\omega = d \eta \, ,
\ee
where $\varphi^*$ is the pullback. The symmetry-breaking parameters now take value in the quotient $G/G_0$. If $G$ is simple, then $G_0$ must be discrete, otherwise we can have continuous anomaly-free subgroups, e.g. for $G= U(1) \times U(1)$. 
Relevant examples include the $U(1)_A U(1)_V$ and $U(1)_A U(1)_V^2$ mixed anomalies in (1+1)d and (3+1)d and the preservation of discrete subgroups of $G$. These are often interesting, as $G/G_0$ inherits nontrivial topology from the quotient procedure.

\begin{figure}[t]
    \centering
    \begin{tikzpicture}
     \draw[fill=white!70!gray] (3,1)--(0,2) --(0,0) -- (3,-1) ;   
     \draw (0,0) -- (4,0);   \draw (0,2) -- (4,2);  

      \begin{scope}[shift={(3,-1)}]

      \def\N{100}        
  \def\A{0.045}        
  \def\freq{3}       

  \fill[white!70!cyan, opacity=0.5]
    plot[domain=0:1, samples=\N, variable=\t]
      ({\A*sin(360*\freq*\t)}, {2*\t}) --       
    plot[domain=1:0, samples=\N, variable=\t]
      ({4 + \A*sin(360*\freq*\t)}, { 2*\t}) -- 
    cycle;

    \draw[line width=3, color=cyan]  
     plot[domain=0:1, samples=\N, variable=\t]
      ({\A*sin(360*\freq*\t)}, {2*\t});

   \draw[cyan] (0,0) -- (4,0); \draw[cyan] (0,2) -- (4,2);

      \foreach \i in {0,...,\N} {
    \pgfmathsetmacro{\t}{\i/\N}
    \pgfmathsetmacro{\xleft}{\A*sin(360*\freq*\t)}
    \pgfmathsetmacro{\xright}{4 + \A*sin(360*\freq*\t)}
    \pgfmathsetmacro{\y}{2*\t}
    \pgfmathsetmacro{\shadeval}{abs(sin(360*\freq*\t))}
    \pgfmathsetmacro{\opacity}{0.05 + 0.2*\shadeval}
    \draw[cyan!50!black, line width=0.5pt, opacity=\opacity]
      (\xleft,\y) -- (\xright,{ \y});
  }
    \end{scope}
    \node at (4,1.5) {$\omega(A)$};
    \node at (5,0) {$\gamma(A,g)$};
    \node[below] at (3,-1) {$B_g$};
     
    \end{tikzpicture}
    \caption{An upgraded inflow picture including a (modulated) boundary condition. Maintaining bulk topological invariance requires decoration by a $G$-enriched topological action $\gamma(A,g)$. The dynamical physical modes live on the gray area of the figure.}
    \label{fig: inflow}
\end{figure}

\section{Examples and Applications} We now give several examples of our construction together with some applications.

\subsection{(1+1) dimensions} Our first example concerns anomalies in 1+1 dimensions. For a non-abelian group $G$ with gauge field $A$, the anomaly is encoded in the non-abelian Chern-Simons term:
\be
\omega(A) = \frac{\chi}{4\pi} \Tr\left[ A d A  + \frac{2}{3} A^3 \right] \, ,
\ee
transgression gives the minimally coupled WZW term as $\gamma(A,g)$:
\be
\gamma(A,g) = \frac{\chi}{4 \pi} \Tr  dg g^{-1} \, A + \chi \Gamma_{WZW}(g) \, ,
\ee
where
\be
\Gamma_{WZW} = \frac{1}{12 \pi} \Tr (g^{-1} d g)^3  = \frac{1}{2 \pi} H_{abc}(\sigma) d \sigma^a d \sigma^b d \sigma^c \, ,
\ee
is the standard WZW term for $G$. In the boundary formula we choose a trivialization $H = d B$. The effect of the WZ term on defect conformal manifolds has already been noted in \cite{Choi:2025ebk} in the context of boundary conditions for (1+1)d WZW models. The current derivation shows that it universally depends on the bulk 't Hooft anomaly.
The above is a minimal choice for $\gamma$. If $G$ is simply connected there are no pure $G$ SPTs and the NLSM does not admit any nontrivial discrete theta angle. Thus there are no defect anomalies to be considered in this case.

If $G$ is not simply connected the story is richer. For $G = U(1)$ the WZW term is trivial, however we still have an interesting contribution to $\gamma$:
\be
\gamma(A,g)^{U(1)} = \frac{\chi}{4 \pi} \sigma \, d A \, .
\ee
For fixed $\sigma$, this is a $U(1)$ SPT (for a bosonic theory $\chi$ must be even, so we recover the known periodicity) We will show later that this probes the nontrivial $\pi_1$ of the defect conformal manifold by a phenomenon akin to the Thouless pump. A similar formula holds if $G = U(1) \times U(1)$ has a mixed 't Hooft anomaly:
\be
\gamma(A_1, g_2)^{U(1) \times U(1)} = \frac{\chi}{2 \pi} \sigma_2 d A_1 \, .
\ee
If instead $G$ is a central quotient of a simply connected group, we can turn on a flat $B$ field proportional to the second Stiefel-Whitney class of the $G$ bundle $B \sim \omega_2(BG)$. This is the classic example of a defect anomaly. We will now outline several applications.

\noindent \textbf{Boundary Thouless pump.} Consider a boundary condition $B_\sigma$ that breaks a bulk anomalous $U(1)$ symmetry. This occurs, for example, in the compact boson, where the diagonal $U(1)^d \subset U(1)^m \times U(1)^w$ symmetry has an 't Hooft anomaly with coefficient $\chi = 1$.

For a non-modulated boundary condition, we define the defect charge density via the response:
\be \label{eq: pump}
Q_B(\sigma) = \frac{\delta}{\delta A} F_{B} = \chi \frac{\sigma}{2\pi} \, .
\ee
The point $\sigma = 0$ is usually distinguished by a further discrete symmetry (e.g. time-reversal or charge conjugation). As $\sigma$ is slowly varied, the accumulated charge changes continuously. Upon traversing a closed, non-contractible loop in the boundary conformal manifold, we obtain:
\be
Q_{B}(\sigma + 2 \pi) - Q_B(\sigma) = \chi \, ,
\ee
indicating that $\chi$ units of charge accumulate on the defect. Equivalently, this effect can be described as the stacking of a nontrivial invertible $U(1)$-symmetric TQFT, captured by an integer 1d Chern–Simons term:
\be
\exp\left(i \chi \int_{B} A \right) \, .
\ee
This phenomenon can be interpreted as an anomaly in the space of \emph{defect} couplings, which is sourced by the bulk anomaly.

As an example, consider a free compact scalar field with radius $R$—which has $\chi = 1$—and impose a Neumann boundary condition $|N, \sigma\rangle$\footnote{The Dirichlet boundary condition can be treated in the same manner after $T$-duality.}, described by the action \cite{Choi:2023xjw}:
\be
S = \frac{R^2}{4\pi}\int d X \wedge \star d X + \frac{\sigma}{2\pi} \int_{B} d X \, .
\ee
The anomalous symmetry is the diagonal $U(1)^d \subset U(1)^m \times U(1)^w$. This theory can be coupled to a $U(1)^d$ background gauge field $A$ as:
\be
S[A] = \frac{R}{4\pi}\int (dX - A) \wedge \star d X + i\frac{\sigma}{2\pi} \int_{B} (d X - A_{B}) \, .
\ee
Taking the functional derivative with respect to $A_{B}$ reproduces equation~\eqref{eq: pump} with $\chi = 1$, as expected.

\vspace{2mm} \noindent \textbf{Consequences of the WZW term.} In the non-abelian case there is in general no SPT pumping induced by the bulk 't Hooft anomaly. On the other hand, we must take into account the effects of the non-abelian WZW term instead.

On a closed (1+1)d manifold, the WZW term is well-defined, however, once we include a boundary, we must be careful in its description. 
In the usual treatment of boundary conditions for WZW models, the WZW field $g$ is restricted to a conjugacy class $N \subset G$ on which the WZW term globally trivializes: $H \vert_N = d B$\cite{Alekseev:1998mc}. The quantity:
\be
\int H - \int_{D} B \, ,
\ee
where $D$ is a disk enclosing the boundary, is then well-defined. This condition ensures that the B field is globally well-defined. For our applications, instead we will only choose a local trivialization $H = d B$. This is sufficient to discuss small modulations. The effect of large modulations (i.e. global topological data), which we expect to also be important, will be left for future study.

To discuss physical consequences we set the background gauge field $A=0$ and expand around a fixed boundary condition $\sigma^a$ with a small modulation $\delta \, \sigma^a$. At second order in the small modulation the leading contribution is localized on the physical boundary and equals:
\be
\gamma(0,g)^{(2)} \sim \frac{\chi}{2\pi}\int_{\text{B}} B_{ab}(\sigma) \delta \sigma^a d \delta \sigma^b \, .
\ee
Differentiating with respect to $\delta \sigma^a$ twice we learn that the tilt OPE must take the form:
\be
t^a(\bx) \, t^b(\by) \sim \left[ \frac{G_{ab}}{(\bx - \by)^2} + i \chi \frac{B_{ab}}{2\pi} \partial_\bx \delta(\bx - \by) \right] \unit + ...
\ee
this a contact term enters the algebra of the modes $t_n^a =\int_0^{2 \pi} d \theta \, e^{i n \theta} t^a(\theta)$ of the tilt operator on the circle via:
\be
t_n^a \,  t_m^b  = n \chi \delta_{n + m,0} \left(\delta_{ab} + i B_{ab} \right) + . . .  \, ,
\ee
where the first term is fixed by the bulk Kac-Moody symmetry $g_{ab} = \chi \delta_{ab}$. Notice that this is the natural complexified coupling between metric and B field in a sigma model.

At third order, we can instead consider a variation of the form $\delta \sigma = \delta \lambda + \delta \sigma'$, with $\delta \lambda$ constant. The leading term in $\gamma(0,g)$ then is:
\be
\gamma(0,g)^{(3)} \sim \frac{\chi}{2\pi} \int_{\text{bdry}} H_{abc}(\sigma) \delta \lambda^a \delta {\sigma'}^b d \delta {\sigma'}^c \, . 
\ee
This leads to the following contact term in the tilt three-point function:
\be \label{eq: contect3p}
\bigg\langle \int d \bx t^a(\bx) \, t^b(\by) t^c(\bz)  \bigg\rangle \supset i \frac{\chi H^{abc}(\sigma)}{\pi} \partial_\bz \delta(\by - \bz) \, .
\ee
Conformal symmetry requires that, at separated points, $\langle t^a(\bx) t^b(\by) t^c(\bz) \rangle = f^{abc}/ (\bx - \by) (\bx - \bz) (\by - \bz)$ with $f^{abc}$ completely antisymmetric (this avoid an anomalous dimension in conformal perturbation theory). We can integrate this against a test function $f(\bz)$ (for simplicity on the circle). Focusing on the singularities arising from coincident points shows that $f_{abc} = - (2/\pi)^3  \chi H_{abc}$ is fixed by the bulk anomaly $\chi$. The relation between the WZW term and the tilt 3pf coefficient was pointed out in \cite{Drukker:2025dfm}, here we just remark that its physical origin is the matching of the bulk 't Hooft anomaly.

To conclude, notice that even in the absence of bulk 't Hooft anomalies a flat B-field is allowed. This reflects defect anomalies and leads to nontrivial consequences for the algebra of tilt operators. 

\subsection{(2+1) dimensions}
In (2+1)d there are no continuous (infinite order) ´t Hooft anomalies. However there are interesting mixed anomalies between flavor symmetries/ diffeomorphisms and spacetime parity. 

The statement of the parity anomaly is that the system can be realized in an anomaly-free manner at the boundary of a P-invariant topological insulator:
\be
\omega(A, R) = \pi \int \left( \Hat{A}(R) + \frac{F}{2\pi}^2 \right) \, ,
\ee
Where we have taken $G=U(1)$ for concreteness. On open manifolds, this (naively) becomes a Chern-Simons term with ill-quantized level $k = 1/2 + \bZ$. The same is true for the gravitational term $\hat{A}(R)$, which describes an ill-quantized gravitational Chern-Simons term.
The topological phase is only parity-invariant on open manifolds up to the addition of a boundary (integer-quantized) Chern-Simons term: this originates the anomaly. Boundary conditions then must either break parity of the flavor symmetry. 

\vspace{2mm} \noindent \textbf{Parity-breaking b.c.} We first examine the parity-breaking boundary conditions. These for example arise naturally in the theory of a single Dirac fermion $\psi$ and read:
\be
\gamma^\perp \psi = \pm \psi \, .
\ee
A parity transformation flips the sign of $\gamma^\perp$, exchanging the two boundary conditions. The boundary theory can host chiral degrees of freedom by the anomaly.
The anomaly implies that the chiral central charges $c_\pm$ and levels $k_\pm$ for the two boundary conditions differ by one unit:
\be
c_+ - c_- = 1/2 \, \ \ \  k_+ - k_- = 1 \, .
\ee
In the free fermion example, this can be seen by considering the system on a slab, with the boundary conditions on the two ends obtained by turning on a large positive/negative mass on the two sides. Taking the slab to be very short we end up with the theory of a single Weyl fermion, which saturates the above predictions.

\vspace{2mm} \noindent \textbf{Parity-preserving b.c.} Closer to our discussion is the case of parity-preserving boundary conditions. In the free-fermion examples these can be described by splitting the Dirac field into Majorana modes $\psi = w_+ + i w_-$ and imposing the boundary conditions:
\be
\gamma^\perp w_\pm = \pm w_\pm \, .
\ee
Parity exchanges the $\pm$ boundary conditions, but this can be undone by an $O(2)$ transformation. However the continuous $U(1)$ symmetry rotating the Majoranas is broken down to fermion parity. 
Let us concentrate on the effect of the $U(1)$ part of the anomaly. The ill-quantized Chern-Simons term can be made sensible by the addition of the modulated field $\sigma$, becoming:
\be
\frac{1}{8 \pi} \int (A - d \sigma) F \, ,
\ee
which is exactly gauge invariant. To obtain this we have added a boundary SPT:
\be
\gamma(A,\sigma) = \pi \int \frac{d \sigma}{2 \pi} \, \frac{F}{2 \pi} \, .
\ee
Adiabatically changing $\sigma \to \sigma + 2 \pi$ we obtain the pumping of a $\theta = \pi$ boundary topological angle
$\int_B \frac{F}{4 \pi} $.

\subsection{(3+1) dimensions} In (3+1)d, the anomaly is slightly more intricate. It takes the form:\footnote{Our conventions are such that a single Weyl fermion contributes $\chi=1$.}
\be
\alpha(\lambda,A)= \frac{\chi}{24 \pi^2} \int \Tr d \lambda \left(A d A - \frac{2}{3}A^3 \right) \, ,
\ee
which leads to:
\be
\text{WZ}[A,\lambda_1, \lambda_2]  =
\frac{\chi}{(2\pi)^2} \int_{B} \Tr (\lambda_1 d \lambda_2 - \lambda_2 d \lambda_1) F  \, .
\ee
This is derived from the (4+1)d CS action:
\be
\omega(A) = \frac{\chi}{24 \pi^2} \int \Tr \left( A F^2 - \frac{1}{2} A^3 F + \frac{1}{10} A^5 \right) \, .
\ee
Again transgression gives:
\be
\gamma(A,g) = \chi \Gamma_{WZW}^{(3+1)}(g) + ...
\ee
where the dots indicate higher order terms in the gauge field to ensure gauge invariance. Notice that contrary to the $(1+1)d$ case, symmetry-breaking boundary conditions can host a variety of defect anomalies, even for simple Lie groups. In this short paper, we will only discuss a simple application to the (3+1)-dimensional setting.

\vspace{2mm} \noindent \textbf{Boundary Quantum Hall effect.} For abelian symmetries an analogous SPT pumping effect takes place on the boundary. In this case we have two anomalies, a pure $U(1)^3$ anomaly $\chi$ and a mixed $U(1)$-gravitational anomaly $\chi_g$:\footnote{Explicitly, $\Hat{A}(R) = \frac{1}{48(2\pi^2)} \Tr R \wedge R = d \text{CS}_g$.}
\be
\omega(A, R) = \frac{\chi}{24\pi^2} \int A dA^2  + 2 \pi \chi_g \int A \wedge  \Hat{A}(R)\, ,
\ee
leading to:
\be
\gamma(A,R,g) = \frac{\chi}{24\pi^2} \int d \sigma \wedge A dA + 2 \pi \chi_g \int d \sigma \wedge \text{CS}_g \, .
\ee
In this case this leads to a boundary current:
\be
\star \mathcal{J}_{B} = \begin{cases} \ds \chi \frac{\sigma}{12 \pi^2} F \, , \quad  &U(1)^3 \ \text{anomaly} \\[1em]
\ds \chi \frac{\sigma}{2 \pi^2} F \, , \quad &U(1)_A U(1)_V^2 \ \text{anomaly} \, .
\end{cases}
\ee
Note that the Hall conductance $\ds \sigma_H \sim \chi \frac{\sigma}{\pi}$ can be irrational. We predict such a current to arise, for instance, at the boundary of a free Dirac fermion system in the presence of an external electric field. 

Focusing on the mixed anomaly case, we observe that adiabatic variation of $\sigma$ along a closed loop in coupling space results in the stacking of the boundary with a $U(1)_\chi$ SPT:\footnote{To be precise, since fermion parity is unbroken on the boundary, a closed loop on the conformal manifold corresponds to $\sigma \to \sigma + \pi$.}
\be
\exp \left( i \frac{ \chi}{4 \pi} \int_{B} A d A + 2 \pi i \chi_g \int_B \text{CS}_g \right)  \, .
\ee
Again, we interpret this as an anomaly in the space of defect couplings. This leads to the conclusion that a bulk 't Hooft anomaly manifests as an anomaly in the space of couplings on the symmetry-breaking defect.

As an example, consider a boundary condition for a single Dirac fermion $\psi = (\chi_L, \, \chi_R)$.\footnote{We focus here on the mixed anomaly case, corresponding to $\chi=1$.} The Dirac fermion has a mixed axial anomaly with $\chi=\chi_g =1$.
In (3+1)d, standard (vector-preserving) boundary conditions for the Dirac fermion relate the left- and right-moving Weyl components as:
\be \label{eq: BCfermions}
B_\sigma: \ \ \ \ \chi_L^{\alpha} = e^{2 i \sigma} {(\sigma^\perp)^\alpha}_{\dot{\alpha}} \, \chi_R^{\dot{\alpha}} \, ,
\ee
thus explicitly breaking the (anomalous) axial symmetry—which acts by shifting $\sigma$—to the $\bZ_2$ subgroup corresponding to fermion parity $(-)^F$. The boundary conformal manifold is $U(1)/\bZ_2$, so that $\sigma \simeq \sigma + \pi$.

These boundary conditions can be realized via a half-space RG flow, where a complexified mass term is introduced:
\be
 |M| \left( e^{2 i \sigma} \chi_L^\dagger \sigma^\perp \chi_R + e^{-2 i \sigma} \chi_R^\dagger \sigma^\perp \chi_L \right) \Theta(x_\perp)
\ee
see Figure \ref{fig: halfspaceRG}. The theory in the right half-space is trivially gapped, but in the presence of background fields for the $U(1)_V$ symmetry, it becomes a nontrivial SPT:
\be \label{eq: nuSPT}
\nu_\sigma(F_V) = \sigma \int_{+}\left(\frac{F_V}{2\pi}\right)^2 + 2 \pi \sigma \int_{+} \Hat{A}(R)\, .
\ee
To derive this, one observes that for real positive mass UV counterterms can be chosen so that the SPT is trivial. A complex mass $|M| e^{2 i \sigma}$ can be rotated into a positive one by a constant axial transformation. The anomaly then produces the SPT $\nu_\sigma$, which is consistent with the fact that $(-)^F$ remains unbroken. Moreover, the boundary condition \eqref{eq: BCfermions} immediately follows by setting the mass deformation to zero at the boundary. The result matches the inflow action we have previously obtained.
As $\sigma$ winds around the boundary conformal manifold, the response pumps a $U(1)_1$ integer Quantum Hall state upon encircling the conformal manifold.

If we promote the $A_V$ gauge field to a dynamical one, the axial symmetry is known to become a non-invertible $\bQ/\bZ$ symmetry \cite{Choi:2022jqy,Cordova:2022ieu}. It would be interesting to explore how the structure of boundary conditions in QED is affected by this non-invertible structure.

\begin{figure}
    \centering
    \begin{tikzpicture}
    \shade[left color=white,  right color =white!10!cyan] (0,0) -- (1.5,0) -- (1.5,2) -- (0,2) -- cycle; 
    \shade[left color=white!50!cyan, right color=white] (1.5,0) -- (3,0) -- (3,2) -- (1.5,2) -- cycle; 
    \node at (2.25,1.75) {$|M| e^{2 i \sigma}$};
    \draw[thick,cyan,dashed] (1.5,0) -- (1.5,2);
    \draw[blue] plot[domain=0:1, variable=\t] ({3*\t},{0.75+3/360*atan(360*(\t-0.5))});
    \node at (4,1) {\LARGE $\leadsto$};
    \begin{scope}[shift={(5,0)}]
       \shade[left color=white,  right color =white!10!cyan] (0,0) -- (1.5,0) -- (1.5,2) -- (0,2) -- cycle;    
       \draw[very thick] (1.5,0) node[below] {$B_\sigma$} -- (1.5,2);
       \node at (2.25,1) {$\nu_\sigma$};
    \end{scope}
    \end{tikzpicture}
    \caption{Half-space RG flow giving rise to a $U(1)_V$-symmetric boundary condition $B_\sigma$ for a single Dirac fermion. This realizes an interface between the massless Dirac fermion and the topological insulator $\nu_\sigma$.}
    \label{fig: halfspaceRG}
\end{figure}

\section{Defect Anomalies Matched by Defect Couplings}
To conclude, let us consider an alternative scenario: a defect system with no bulk 't Hooft anomaly, but a nontrivial \emph{defect anomaly}.\footnote{I thank Andrea Antinucci for discussions on this point.} The notion of a defect anomaly was introduced in \cite{Antinucci:2024izg}, generalizing familiar constructions such as line defects carrying fractionalized representations of the flavor symmetry \cite{Delmastro:2022pfo,Brennan:2022tyl}. Much like bulk 't Hooft anomalies, a defect anomaly can be encoded via inflow from a $(p+1)$-dimensional SPT:
\be
\nu \left(A, \mathbf{B} \right) \in \text{SPT}_{G \times \mathbf{K}}^{p+1} \, ,
\ee
where $A$ and $\mathbf{B}$ denote background gauge fields for the bulk (G) and defect ($\mathbf{K}$) symmetries, respectively. We think of its inflow theory as realized on a $(p+1)$-dimensional bulk surface terminating on the defect, as shown in Figure~\ref{fig: defectanom}.

\begin{figure}
    \centering
        \begin{tikzpicture}[baseline={(-3,0)}]
     \def\N{100}              
  \def\A{0.1}             
  \def\L{2.0}              
  \def\freq{2}            

  \shade[left color=white, right color=red!50!white]
    (0,0)
    -- plot[domain=0:\L, samples=\N, variable=\y]
         ({\A*sin(360*\freq*\y/\L)}, \y)
    -- (-3,2) -- (-3,0) -- cycle;
\draw[very thick] plot[domain=0:\L, samples=\N, variable=\y]
         ({\A*sin(360*\freq*\y/\L)}, \y);
 \node[below] at (0,0) {$D$};
 \node at (-1.5,1) {$\nu \left(A, \mathbf{B} \right)$};
  \node at (-3.5,1) {(a)};
    \end{tikzpicture} 
    \\[2em]
    \begin{tikzpicture}[baseline={(-3,0)}]
  \def\N{100}      
  \def\A{0.1}     
  \def\L{2.0}      
  \def\freq{2}    
  \def\dx{-3}    

 \shade[shading=axis, shading angle=0,
         left color=white!50!red, right color=red!50!white, middle color=white!90!red]
    plot[domain=0:\L, samples=\N, variable=\y]
      ({\A*sin(360*\freq*\y/\L)}, \y)
    -- ({\A*sin(360*\freq)}, \L) -- ({\dx + \A*sin(360*\freq)}, \L)
    -- plot[domain=\L:0, samples=\N, variable=\y]
      ({\dx + \A*sin(360*\freq*\y/\L)}, \y)
    -- cycle;

    \shade[shading=axis, shading angle=0,
         left color=blue!50!white, right color=white]
    plot[domain=0:\L, samples=\N, variable=\y]
      ({\A*sin(360*\freq*\y/\L)}, \y)
    -- ({\A*sin(360*\freq)}, \L) -- ({-\dx + \A*sin(360*\freq)}, \L)
    -- (-\dx,0) -- cycle;

          \draw[very thick]
    plot[domain=0:\L, samples=\N, variable=\y]
      ({\A*sin(360*\freq*\y/\L)}, \y);
  \draw[very thick]
    plot[domain=0:\L, samples=\N, variable=\y]
      ({\dx + \A*sin(360*\freq*\y/\L)}, \y);    
      \node[below] at (0,0) {$D_{\sigma}$};
      \node[below] at (-3,0) {$D_{\text{UV}}$};

      \node at (-1.5,1) {$\nu \left(A, \mathbf{B} \right)$};
      \node at (1.5,1) {$\nu \left(\omega, \mathbf{B} \right)$}; 
      \node at (-3.5,1) {(b)};
    \end{tikzpicture}
    \caption{Pictorial representation of defect anomaly inflow. (a) An anomalous defect $D$ dressed by an invertible theory $\nu$. (b) A defect RG flow to a symmetry-breaking defect $D_\sigma$, where the defect anomaly is matched by an anomaly in the space of defect couplings.}
    \label{fig: defectanom}
\end{figure}
Unlike bulk 't Hooft anomalies, defect anomalies do not require symmetry breaking on $D$: they can be saturated by symmetric defect QFTs that share the same anomaly. However, one may envision a defect RG flow from a symmetric defect $D_{\text{UV}}$ to a symmetry-breaking IR defect $D_\sigma$, where as before $\sigma$ parametrizes the breaking of a bulk symmetry. This scenario can be forced, for instance, by turning on a relevant defect deformation in a nontrivial G representation.

The defect anomaly must still be matched and this occurs through an anomaly in the space of defect couplings, obtained by setting $A = g^{-1} d g$ (a flat connection) in the inflow SPT.

Concrete examples can be constructed using weakly coupled UV defects. Consider, for instance, an ``electric" surface defect in a $U(1)$-symmetric bulk CFT, realized by the coupling of a free compact scalar $X$ of radius $R$:
\be
S = S_{\text{bulk}} + \frac{R^2}{4\pi} \int_{D} dX \wedge \star dX + g\left( e^{i X/R} \Phi^\dagger + \ \text{c.c.} \right) \, ,
\ee
where $\Phi$ is a bulk primary of charge one and dimension $\Delta_\Phi < 2$. By tuning $R$, the deformation can be made weakly relevant, such that the bulk-defect system flows to an interacting fixed point within conformal perturbation theory. The resulting DCFT has a $U(1) \times \mathbf{U(1)}_{w}$ symmetry and a defect anomaly:
\be
\nu(A,\mathbf{A}_w) = \frac{i}{2\pi} \int A \, d \mathbf{A}_w \, .
\ee
This fixed point is generically unstable to U(1)-breaking deformations, as the vertex operator $e^{i X/R}$ has a small anomalous dimension and thus remains relevant. This gives rise to a one-parameter family of deformations:
\be
\lambda \cos(X + \sigma) \, ,
\ee
which defines a $U(1)$-family of symmetry-breaking defects. Minimal coupling in the UV Lagrangian description immediately yields an anomaly in the space of defect couplings:
\be
\nu(\sigma, \mathbf{A}_w) = \frac{i}{2 \pi} \int d \sigma \, d  \mathbf{A}_w \, ,
\ee
upon shifting $X \to X -\sigma$. Such anomaly must be matched by the symmetry-breaking IR defect. The anomaly does not imply that the defect in the IR is gapless, unfortunately. Indeed, in the present case, the defect anomaly can be matched by an appropriate SPT defect:
\be
S_{\text{IR}} = S_{\text{bulk}} + \frac{i \sigma}{2 \pi} \int_{D} d \mathbf{A}_w \, .
\ee

\vspace{2mm} \noindent \textbf{Outlook} In this Letter, we initiated an in-depth exploration of the effects of anomaly-enforced symmetry breaking on boundaries and defects. We described how consistency with a bulk 't Hooft anomaly fixes universal features of the modulated response on the defect. 
Let us conclude by highlighting some promising directions for future investigation:

\begin{enumerate}
    \item \emph{Gravitational and global anomalies.} In this work, we have restricted our attention to perturbative anomalies associated with flavor symmetries. A natural extension is to study the constraints imposed by gravitational anomalies. For instance, it is well known that in (1+1) dimensions, a nonzero chiral central charge forbids the existence of a conformal boundary condition \cite{Jensen:2017eof,Hellerman:2021fla}. It would also be interesting to understand whether gravitational anomalies contribute nontrivially to the modulated effective action for the displacement operator \cite{Gabai:2025zcs}. Likewise, global anomalies for flavor symmetries may induce nontrivial boundary pumping effects, offering insights into the global topology of defect conformal manifolds.

    \item \emph{SymTFT derivation of modulated effective actions.} Several recent works have studied the representation of symmetry on boundary conditions using Symmetry Topological Field Theory (SymTFT) techniques \cite{Huang:2023pyk,Bhardwaj:2024igy,Choi:2024tri,Copetti:2024rqj,Copetti:2024dcz,Cordova:2024iti}. However, these approaches have thus far been limited to discrete symmetries. A SymTFT formulation for continuous symmetries has been recently proposed by several authors \cite{Antinucci:2024zjp,Brennan:2024fgj,Bonetti:2024cjk}. It is natural to expect that the universal terms in the modulated effective action could be derived in full generality within this framework, perhaps along the lines of \cite{Antinucci:2024bcm}.

    \item \emph{Defects and higher-form symmetries.} The formalism developed in this Letter readily generalizes to higher-codimension defects and can be extended to include higher-form symmetries and higher-group structures. While the present discussion has focused primarily on boundary conditions, we anticipate that these methods will yield nontrivial constraints on defect RG flows across dimensions. We hope to return to this topic in the near future.
\end{enumerate}

\vspace{2mm}
\noindent \textbf{Acknowledgements.} The author thanks Andrea Antinucci, Giovanni Galati, and Giovanni Rizi for extensive collaboration on related projects, and Lorenzo Di Pietro for discussions on defect conformal manifolds, as well as for introducing the author to the concept of ``tilt" operators. The author also thanks Petr Kravchuk for comments and discussions, leading to a revised version of the paper.
The author is supported by STFC grant ST/X000761/1.

 \setlength{\bibsep}{2pt plus 0.3ex}

 \bibliography{mybib.bib}

\end{document}